# Bodge: Python package for efficient tight-binding modeling of superconducting nanostructures

Jabir Ali Ouassou [1,2]

**1** Department of Computer Science, Electrical Engineering and Mathematical Sciences, Western Norway University of Applied Sciences, NO-5528 Haugesund, Norway **2** Center for Quantum Spintronics, Department of Physics, Norwegian University of Science and Technology, NO-7491 Trondheim, Norway

## Summary

Bodge is a Python package for constructing *large-scale real-space tight-binding models* for calculations in condensed matter physics. "Large-scale" means that it should remain performant even for lattices with millions of atoms, and "real-space" means that the model is formulated in terms of individual lattice sites and not in momentum space, for example.

Although general tight-binding models can be constructed with this package, the main focus is on the Bogoliubov–De Gennes ("BoDGe") Hamiltonian used to model superconductivity in the clean limit (de Gennes, 1966; Zhu, 2016). The package is designed to be easy to use, flexible, and extensible—and very few lines of code are required to model heterostructures containing, e.g., conventional and unconventional superconductors, ferromagnets and antiferromagnets, altermagnetism, and spin-orbit coupling.

In other words: If you want a lattice model for superconducting nanostructures, and want something that is computationally efficient yet easy to use, Bodge should be a good choice.

## Statement of need

In condensed matter physics, a standard methodology for modeling materials is the *tight-binding model*. In the context of electronic systems (e.g., metals), the electrons in such a model typically "live" at one atomic site, but from time to time "hop" over to neighboring atoms. By including a spin structure as well in this formalism—meaning that we keep track of what spins each electron has, and whether the spins "flip" during various interactions that are permitted on this lattice—we can model a wide variety of physical phenomena including superconductivity and magnetism. Mathematically, this is often expressed in the language of quantum field theory: We define one operator $c_{i\sigma}^{\dagger}$ that "puts" an electron with spin $\sigma \in \{\uparrow, \downarrow\}$ on an atomic site with some index $i$, and another operator $c_{i\sigma}$ that "removes" a corresponding electron. The Hamiltonian operator $\mathcal{H}$ of the system is then constructed out of these electron operators—and this can in turn be used to calculate, e.g., the ground-state energy, electric currents, superconducting order parameters, and other relevant material properties.

To do anything useful with that Hamiltonian *on a computer*, however, you typically have to translate it to a matrix form. This is where Bodge enters the picture:

- It provides an easy-to-use Pythonic interface for constructing the Hamiltonian of a tight-binding system. Particular focus has been placed on making it easy to describe systems that include various forms of superconductivity and magnetism, making it a great choice for modeling, e.g., superconductivity in magnetic heterostructures.
- It scales well to large systems. For efficiency, it uses SciPy sparse matrices internally (Virtanen et al., 2020), and it constructs large Hamiltonians in $\mathcal{O}(N)$ time and memory where $N$ is the number of sites. According to [my benchmarks](#), the performance is similar to [Kwant](#) ([Groth et al., 2014](#)), which is the state of the art for numerical condensed matter physics. The results can be returned in most NumPy or SciPy matrix formats.

- It is designed to be extensible. For instance, while Bodge currently only implements square and cubic lattices (via the `CubicLattice` class), it can be used to construct Hamiltonians on triangular or hexagonal lattices if you want: you just need to create your own subclass of the `Lattice` base class and implement two-to-three short iterators that describe how to iterate through your lattice. (Specifically: the methods `.sites`, `.bonds`, and `.edges` need separate implementations per `Lattice` type.)
- Some convenience methods are provided to help you with the next steps of your calculations: Extracting the local density of states (LDOS), calculating the free energy, diagonalizing the Hamiltonian, etc. (Some more advanced algorithms live on the `develop` branch on GitHub, but have not yet been assimilated into the official package.)
- The code itself follows modern software development practices: Full test coverage with continuous integration (via `pytest`), fast runtime type checking (via `beartype`), and PEP-8 compliance (via `black`).

There are two main alternatives that arguably fill a similar niche to Bodge: Kwant ([Groth et al., 2014](#)) and Pybinding ([Moldovan et al., 2020](#)). Compared to these packages, the main benefit of Bodge is the focus on the BdG Hamiltonian in particular. For instance, using Kwant, it is up to the user to declare that each lattice site has four degrees of freedom (spin-up electrons, spin-down electrons, spin-up holes, and spin-down holes), and to ensure that you construct a Hamiltonian with the correct particle-hole symmetries. Bodge, however, *assumes* that these are the only relevant degrees of freedom, and enforces the relevant symmetries by default. In practice, this means that Kwant can be used to study a broader variety of physical systems, whereas Bodge can provide a friendlier syntax for users who work specifically on superconducting systems. Both packages support both NumPy arrays and SciPy sparse matrices as output formats, and both provide similar performance in the limit of large systems.

Sparse matrices, including, e.g., the *Compressed Sparse Row* (CSR) format used below, have the advantage that they only store the non-zero elements of a matrix. For a typical tight-binding model with nearest-neighbor hopping terms, the Hamiltonian matrix that describes a lattice with $N$ atoms has $\mathcal{O}(N^2)$ elements where only $\mathcal{O}(N)$ are non-zero. Thus, algorithms that leverage sparse matrices often result in at least an $\mathcal{O}(N)$ reduction in CPU and RAM requirements, which becomes highly significant for large systems. Even larger performance enhancements can be obtained in some limits ([Nagai, 2020](#); [Weiße et al., 2006](#)), although it depends on your system whether the required approximations provide a good trade-off between accuracy and speed. However, *dense matrices* (i.e., NumPy arrays) allow for simpler solution algorithms based on, e.g., full matrix diagonalization, and can become faster than sparse matrix algorithms for small systems or when leveraging GPU acceleration. Notably, the computational complexity of sparse matrix algorithms often come with a large constant prefactor (e.g., the order of a Chebyshev matrix expansion), which can actually result in worse performance for smaller systems. For this reason, both sparse and dense matrices are fully supported by Bodge, allowing the user to pick the most suitable matrix format for the task at hand.

## Examples and workflows

Introductory examples of how to use Bodge are provided in the [official documentation](#). Examples of research problems that have been studied using Bodge include superconductor/altermagnet heterostructures ([Ouassou et al., 2023](#)) and RKKY interactions in unconventional superconductors ([Ouassou et al., 2024b](#), [2024a](#)). These papers also describe some sparse matrix algorithms that can be used together with Bodge, including Chebyshev expansion of the Fermi matrix ([Benfenati, 2022](#); [Goedecker & Colombo, 1994](#); [Ouassou et al., 2023](#); [Weiße et al., 2006](#)) and a quick way to calculate the local density of states ([Nagai et al., 2017](#); [Ouassou et al., 2024b](#)).

For a simple example of how this package can be used, consider a $64a \times 64a$ square lattice.

Let's assume that the whole system is a metal with chemical potential $\mu = 1.5t$, where $t = 1$ is the hopping amplitude. Moreover, let's assume that half the system ($x < 32a$) is a conventional $s$-wave superconductor with an order parameter $\Delta_s = 0.1t$, whereas the other half is a ferromagnet with exchange field $\mathbf{M} = M_z \mathbf{e}_z$. To describe the Hamiltonian corresponding to this system we can use the following code:

```python
from bodge import *

# Tight-binding parameters
t = 1
μ = 1.5 * t
Δs = 0.1 * t
Mz = 0.5 * Δs

# Construct the Hamiltonian
lattice = CubicLattice((64, 64, 1))
system = Hamiltonian(lattice)

with system as (H, Δ):
    for i, j in lattice.bonds():
        H[i, j] = -t * σ0
    for i in lattice.sites():
        if i[0] < 32:
            H[i, i] = -μ * σ0
            Δ[i, i] = -Δs * jσ2
        else:
            H[i, i] = -μ * σ0 - Mz * σ3
```

Note the use of a context manager (`with`-block) to provide an intuitive array syntax for accessing the relevant parts of the Hamiltonian matrix, while abstracting away the underlying sparse matrix details. Afterwards, there are many different ways to use the resulting object.

Some physical observables can be directly calculated using the methods provided in Bodge. For instance, one can use the method `system.ldos(site, energies)` to directly calculate the local density of states at a given lattice site, which can then be used to check for spectral features such as a *superconducting gap* or a *zero-energy peak*. There is also a method `system.free_energy(temperature)` which calculates the *free energy* of the system. By varying parameters in the Hamiltonian (e.g., the orientation of a magnetic field) and then minimizing this free energy, one can determine the ground state of the system, for example.

Most calculations of interest, however, requires that the user implements some code themselves. There are then two main approaches one can take. The classic approach is *matrix diagonalization* which uses dense matrices internally. Bodge provides the method `diagonalize` for this purpose:

```python
E, v = system.diagonalize()
```

The results above contain the positive energies $E_n$ and corresponding state vectors $\mathbf{v}_n$ which satisfy the eigenvalue equation $\mathbf{H}\mathbf{v}_n = E_n \mathbf{v}_n$. (Only positive eigenvalues are returned due to the "Nambu doubling" of degrees of freedom in superconducting systems.) Once the eigenvalues and eigenvectors have been obtained, the user can themselves calculate physical properties of interest from these using equations from standard textbooks on the "Bogoliubov–de Gennes" approach to modeling superconductivity (Zhu, 2016). It is a future goal to incorporate more calculation methods of this kind into the Bodge package itself. Support for matrix diagonalization using GPUs via the CuPy package is also under development.

Examples of physical observables that can be calculated from the eigenvalues and eigenvectors include the superconducting order parameter, electric currents, and spin currents. These can in turn be used to calculate even more material properties. For instance, the critical current of a Josephson junction is defined as the largest electric current that can flow through it for any phase difference, and the critical temperature of a bulk superconductor is defined as the largest temperature at which the superconducting order parameter remains non-zero. These observables can thus be determined via numerical optimization of the electric current and superconducting order parameter, respectively. For instance, the critical temperature can be efficiently determined using a bisection method (Ouassou et al., 2016; Ouassou, 2019).

A modern alternative to matrix diagonalization is a series of algorithms based on Chebyshev expansion of the Hamiltonian matrix (Benfenati, 2022; Covaci et al., 2010; Nagai, 2020; Ouassou et al., 2023; Weiße et al., 2006). These algorithms take advantage of the extreme sparsity of the Hamiltonian matrix, and thus typically provide a significant performance benefit in the limit of very large lattices. One of the main design goals behind Bodge has been to make it trivial to construct sparse representations of the Hamiltonian matrix for this purpose, and it is therefore straight-forward to export the result as, e.g., a CSR sparse matrix:

```
H = system.matrix(format="csr")
```

The user can then easily use the resulting matrix $\mathbf{H}$ to formulate their own sparse matrix algorithms of this kind. For instance, arbitrary matrix functions can typically be evaluated as a Chebyshev expansion $f(\mathbf{H}) = \sum_{m=0}^{M-1} f_m T_m(\mathbf{H})$, where the Chebyshev moments $f_m$ of the function $f(\,\cdot\,)$ are computationally inexpensive to obtain, whereas the Chebyshev matrix polynomials $T_m(\,\cdot\,)$ can be obtained via the recursion relation (Weiße et al., 2006)

$$T_m(\mathbf{H}) = \begin{cases} \mathbf{H}^m & \text{if } 0 \leq m \leq 1, \\ 2\mathbf{H} T_{m-1}(\mathbf{H}) - T_{m-2}(\mathbf{H}) & \text{if } m > 1. \end{cases}$$

The trick is then to find a suitable matrix function $f(\mathbf{H})$ to study your physical system. Using the "Kernel Polynomial Method" (Weiße et al., 2006), the order $M$ of such a Chebyshev expansion can be decreased without serious truncation errors.

## Acknowledgements

I acknowledge very helpful discussions with my PostDoc supervisor Prof. Jacob Linder when learning the BdG formalism, without which the Bodge package would not exist today. I also thank Morten Amundsen, Henning G. Hugdal, and Sol H. Jacobsen for useful discussions on tight-binding modeling in general. Finally, I want to thank Mayeul d'Avezac and Yue-Wen Fang for their constructive input during the referee process, which has improved the Bodge software package and its documentation.

This work was supported by the Research Council of Norway through Grant No. 323766 and its Centres of Excellence funding scheme Grant No. 262633 "QuSpin." During the development of this package, some numerical calculations were performed on resources provided by Sigma2—the National Infrastructure for High Performance Computing and Data Storage in Norway, Project No. NN9577K. The work presented in this paper has also benefited from the Experimental Infrastructure for Exploration of Exascale Computing (eX3), which is financially supported by the Research Council of Norway under contract 270053.